\documentclass[12pt,letterpaper]{article}

\usepackage{units}
\usepackage{cite}

\usepackage[numbers]{natbib}
\usepackage{graphicx}
\usepackage{amsmath}
\usepackage{amssymb}
\usepackage{afterpage}
\usepackage{graphics}
\usepackage{float}
\usepackage{lineno}
\usepackage{rotating}
\usepackage{xspace}
\usepackage{dcolumn}
\usepackage{bm} 

\usepackage[left=1.0in,top=1.0in,bottom=1.0in,right=1.0in]{geometry}

\title{\textbf{Coherent Scattering Investigations at the Spallation Neutron Source: a Snowmass White Paper}}

\def\gtwid{\mathrel{\raise.3ex\hbox{$>$\kern-.75em\lower1ex\hbox{$\sim$}}}}
\def\ltwid{\mathrel{\raise.3ex\hbox{$<$\kern-.75em\lower1ex\hbox{$\sim$}}}}

\newcommand{\gsim}{\mathrel{\rlap{\raisebox{.3ex}{$>$}}
    \raisebox{-.6ex}{$\sim$}}}
\newcommand{\lsim}{\mathrel{\rlap{\raisebox{.3ex}{$<$}}
    \raisebox{-.6ex}{$\sim$}}}

\begin{document}

\maketitle

\begin{center}
D.~Akimov$^{11}$, A.~Bernstein$^{9}$, P.~Barbeau$^{3}$, P.~Barton$^{8}$, A.~Bolozdynya$^{11}$, B.~Cabrera-Palmer$^{19}$, F.~Cavanna$^{23}$, V.~Cianciolo$^{16}$, J.~Collar$^{2}$, R.J.~Cooper$^{6}$, D.~Dean$^{16}$, Y.~Efremenko$^{21,11}$, A.~Etenko$^{11}$, N.~Fields$^{2}$, M.~Foxe$^{18}$, E.~Figueroa-Feliciano$^{12}$, N.~Fomin$^{21}$, F.~Gallmeier$^{16}$, I.~Garishvili$^{21}$, M.~Gerling$^{19}$, M.~Green$^{13}$, G.~Greene$^{21}$, A.~Hatzikoutelis$^{21}$, R.~Henning$^{13}$, R.~Hix$^{16}$, D.~Hogan$^{1}$, D.~Hornback$^{16}$, I.~Jovanovic$^{18}$, T.~Hossbach$^{17}$, E.~Iverson$^{16}$, S.R.~Klein$^{8}$, A.~Khromov$^{11}$, J.~Link$^{22}$, W.~Louis$^{10}$, W.~Lu$^{16}$, C.~Mauger$^{10}$, P.~Marleau$^{19}$, D~Markoff$^{14}$, R.D.~Martin$^{20}$, P.~Mueller$^{16}$, J.~Newby$^{16}$, J.~Orrell$^{17}$, C.~O'Shaughnessy$^{13}$, S.~Penttila$^{16}$, K.~Patton$^{15}$, A.W.~Poon$^{8}$, D.~Radford$^{16}$, D.~Reyna$^{19}$, H.~Ray$^{5}$, K.~Scholberg$^{3}$, V.~Sosnovtsev$^{11}$, R.~Tayloe$^{6}$, K.~Vetter$^{8}$, C.~Virtue$^{7}$, J.~Wilkerson$^{13}$, J.~Yoo$^{4}$, C.H.~Yu$^{16}$\\

\vspace{0.1in}
\small
$^{1}$University of California, Berkeley, Berkeley, CA 94720, USA\\ 
$^{2}$University of Chicago, Enrico Fermi Institute, Chicago, IL 60637, USA\\ 
$^{3}$Duke University, Durham, NC 27708-0754, USA\\ 
$^{4}$Fermi National Accelerator Laboratory, Batavia, IL 60510, USA\\ 
$^{5}$University of Florida, Gainesville, FL 32611, USA\\ 
$^{6}$Indiana University, Bloomington, IN 47405-7105, USA\\ 
$^{7}$Laurentian University, Sudbury, ON P3E 2C6, Canada\\ 
$^{8}$Lawrence Berkeley National Laboratory, Berkeley, CA 94720, USA\\ 
$^{9}$Lawrence Livermore National Laboratory, Livermore, CA 94550, USA\\ 
$^{10}$Los Alamos National Laboratory, Los Alamos, NM 87545, USA\\ 
$^{11}$Moscow Engineering Physics Institute, Moscow, 115409, Russia\\ 
$^{12}$Massachusetts Institute of Technology, Cambridge, MA 02139, USA\\ 
$^{13}$University of North Carolina, Chapel Hill, NC 27599, USA\\ 
$^{14}$North Carolina Central University, Durham NC 27707, USA\\ 
$^{15}$North Carolina State University, Raleigh, NC 27695, USA\\ 
$^{16}$Oak Ridge National Laboratory, Oak Ridge, TN 37831, USA\\ 
$^{17}$Pacific Northwest National Laboratory, Richland, WA 99352, USA\\ 
$^{18}$Pennsylvania State University, University Park, PA 16802, USA\\ 
$^{19}$Sandia National Laboratories, Livermore, CA 94550, USA\\ 
$^{20}$University of South Dakota, Vermillion, SD 57069, USA\\ 
$^{21}$University of Tennessee, Knoxville, TN 37996-1200, USA\\ 
$^{22}$Virginia Polytechnic Institute and State University, Blacksburg, VA 24061, USA\\ 
$^{23}$Yale University, New Haven, CT 06511-8962, USA\\ 

\end{center}

\normalsize
\begin{abstract}
The Spallation Neutron Source (SNS) at Oak Ridge National Laboratory, Tennessee, provides an intense flux of neutrinos in the few tens-of-MeV range, with a sharply-pulsed timing structure that is beneficial for background rejection.  In this white paper, we describe how the SNS source can be used for a measurement of coherent elastic neutrino-nucleus scattering (CENNS), and the physics reach of different phases of such an experimental program (CSI: Coherent Scattering Investigations at the SNS).

\end{abstract}

\section{Introduction}

Coherent neutral-current (NC) neutrino-nucleus scattering (CENNS) was first predicted theoretically in 1974~\cite{Freedman:1977xn} but has never been observed experimentally. The condition of coherence requires sufficiently small momentum transfer to the nucleon so that the waves of off-scattered nucleons in the nucleus are all in phase and contribute coherently. While interactions for neutrino energies in MeV to GeV range have coherent properties, neutrinos with energies less than 50 MeV are most favorable, as they largely fulfill the coherence condition in most target materials with nucleus recoil energy of tens of keV. The elastic NC interaction in particular leaves no observable signature other than low-energy nuclear recoils. Technical difficulties involved in the  development of a large-scale, low-energy-threshold and low-background detectors have hampered the experimental realization of the CENNS measurement for more than three decades. However, recent innovations in dark-matter detector technology (e.g.,~\cite{Aalseth:2012if, Chepel:2012sj})  have made the unseen CENNS reaction testable.  A well-defined neutrino source is the essential component for measurement of CENNS. The energy range of the stopped-pion Spallation Neutron Source (SNS) neutrinos is below 50~MeV, which is the optimal energy to observe pure coherent $\nu$A scattering. 
The detection is within the
reach of the current generation of low-threshold detectors~\cite{Scholberg:2005qs}.  
In addition to being a precise standard-model test, this reaction is also important
for supernova processes and detection.  Physics reach is described in more detail in Section~\ref{coherent_phys} and reference~\cite{Scholberg:2005qs}.

In this document, we describe a program to exploit the SNS neutrinos for CENNS physics, for which the first phase would aim for a first measurement.  Subsequent phases (possibly sharing resources with other initiatives for neutrino physics at the SNS~\cite{Berns:2013usa,Elnimr:2013wfa}) would aim for new tests of the standard model (SM); later phases could probe questions in nuclear physics.  This document draws from a more comprehensive document covering opportunities with neutrinos at the SNS~\cite{Bolozdynya:2012xv}, and discusses recent progress and prospects.

\section{The SNS as a Neutrino Source}\label{source}

The SNS is the world's premier facility for neutron-scattering research, producing pulsed neutron beams with intensities an order of magnitude larger than any currently-operating facility. With the full beam power, $10^{14}$ 1-GeV protons bombard the liquid mercury target in $\sim$700-ns-wide bursts with a frequency of 60 Hz. Neutrons produced in spallation reactions with the mercury thermalize in hydrogenous moderators surrounding the target and are delivered to neutron-scattering instruments in the SNS experiment hall. 

\begin{figure}
\vspace{5mm}
\centering
\includegraphics[width=12cm]{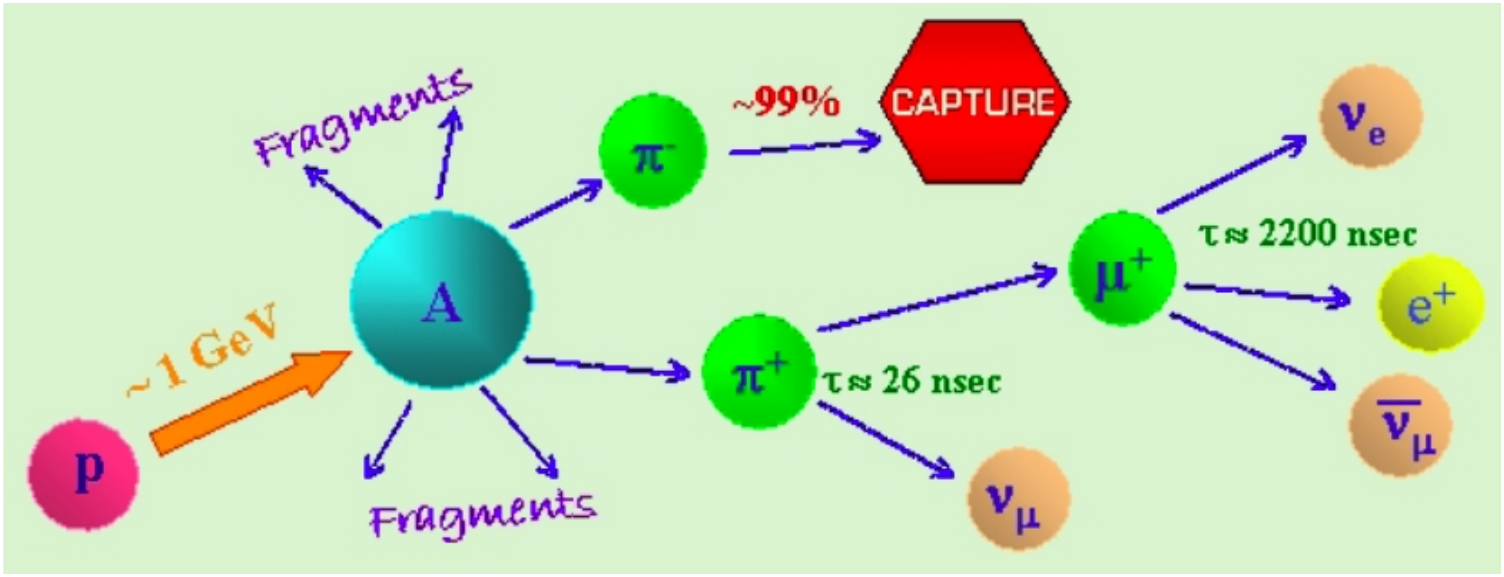}
\vspace{1mm}
\caption{
SNS neutrino production mechanism.
}
\label{fig:snsnu_cartoon}
\end{figure}

As a by-product, the SNS also provides the world's most intense pulsed source of neutrinos in the energy regime of interest for particle and nuclear astrophysics. Interactions of the proton beam in the mercury target produce mesons in addition to neutrons. These stop inside the dense mercury target and their subsequent decay chain, illustrated in Fig.~\ref{fig:snsnu_cartoon}, produces neutrinos with a flux of $\sim 2\times 10^7$ cm$^{-2}$s$^{-1}$ for all flavors at 20 m from the spallation target. This exceeds the neutrino flux at ISIS (where the KARMEN experiment was located~\cite{Zeitnitz:1994kz}) by more than an order of magnitude.

\begin{figure}
\vspace{5mm}
\centering
\includegraphics[width=14cm]{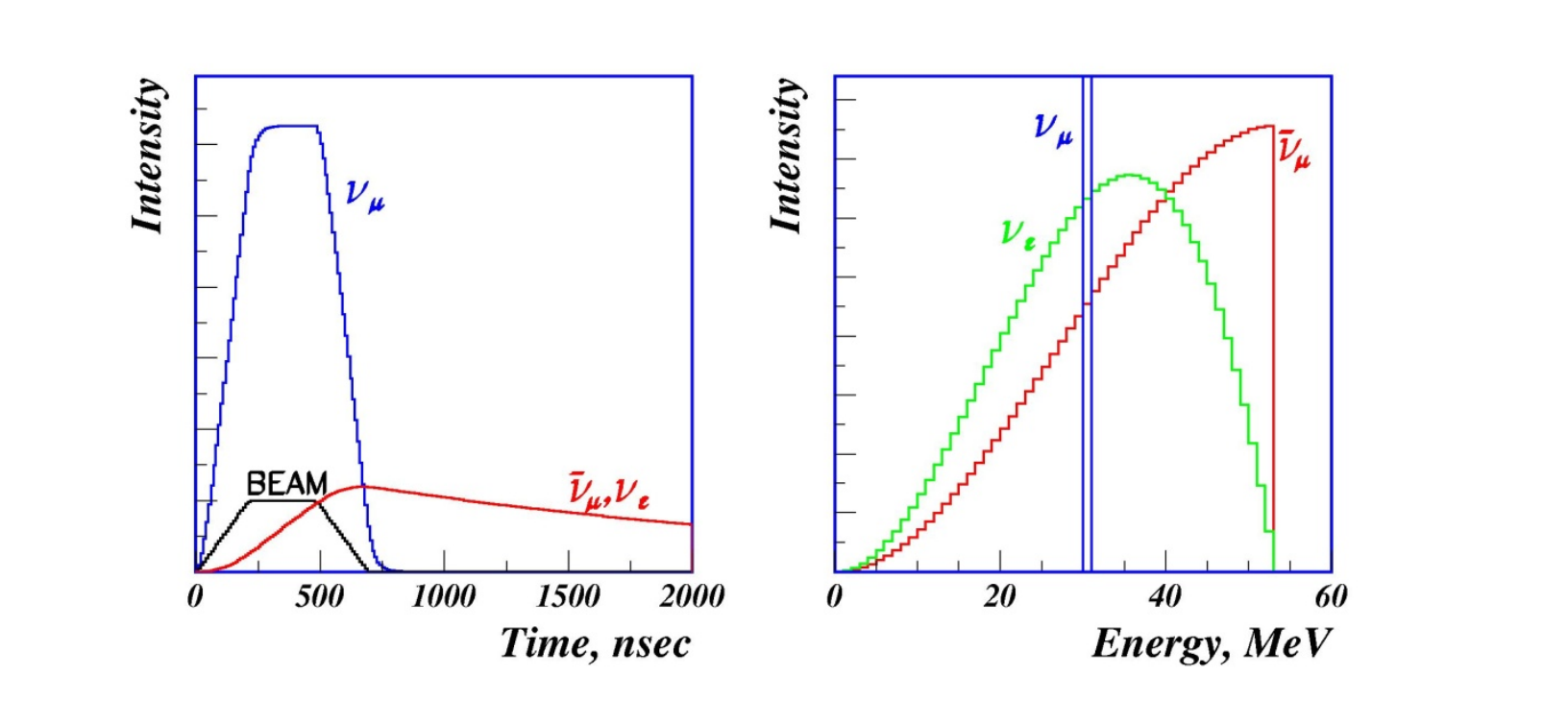}
\includegraphics[width=6cm]{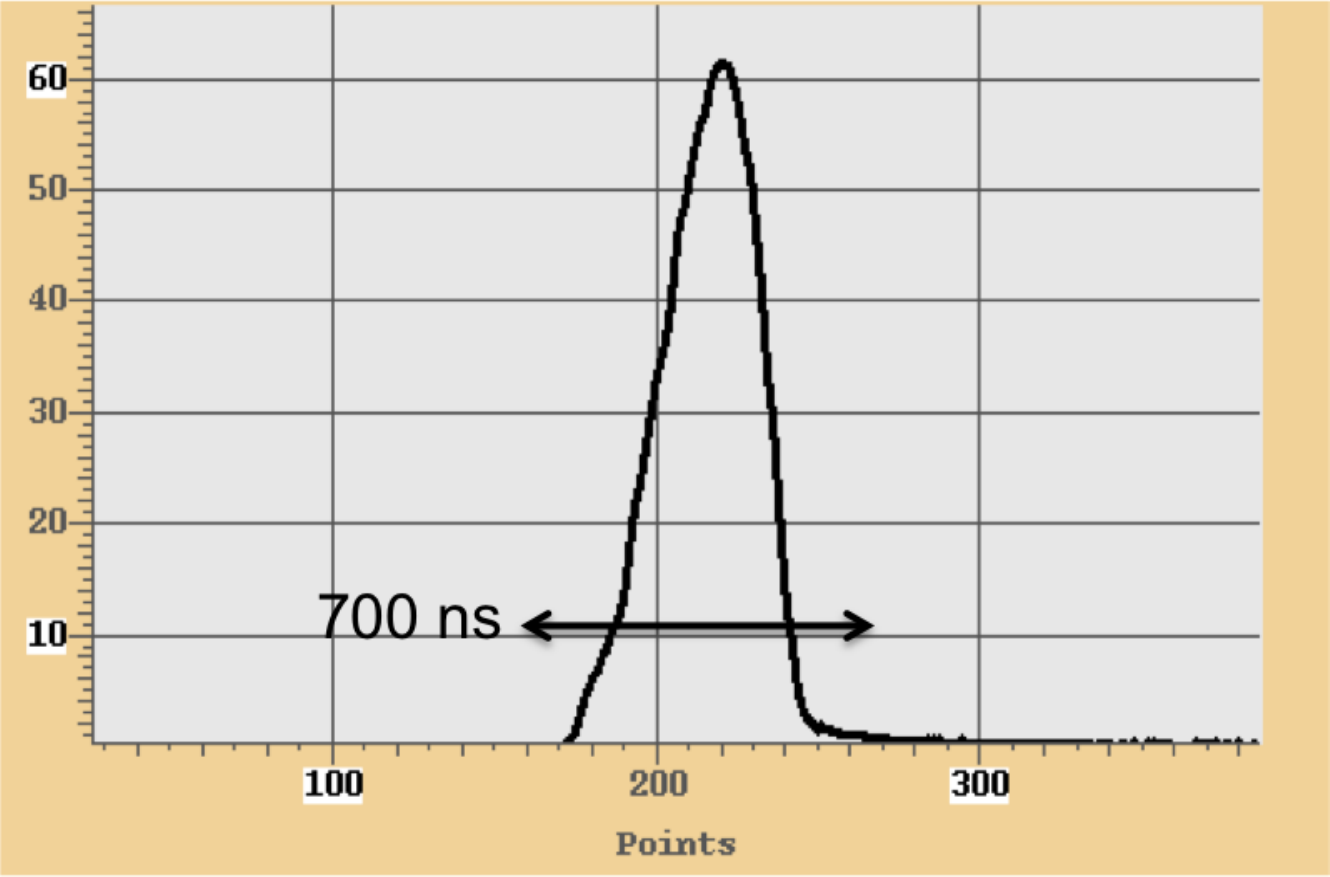}
\vspace{1mm}
\caption{
Time and energy distributions for the different neutrino flavors produced at the SNS.  The top plots are from a 2003 study~\cite{Avignone:2003ep}; the bottom plot shows proton beam structure on target from recent SNS running~\cite{galambos}.  The prompt (pion-decay) component of the neutrino flux should closely follow the proton time structure.
}
\label{fig:sns_nuplots}
\end{figure}

The energy spectra of SNS neutrinos are shown in the right-hand panel of Fig.~\ref{fig:sns_nuplots}. These spectra are well known because almost all neutrinos come from decay-at-rest processes in which the kinematics are well defined. The decay of stopped 
pions produces monoenergetic muon neutrinos at 30~MeV. The subsequent three-body muon decay produces a spectrum of electron neutrinos and muon antineutrinos with energies up to 52.6~MeV.

\begin{figure}
\vspace{5mm}
\centering
\includegraphics[width=12cm]{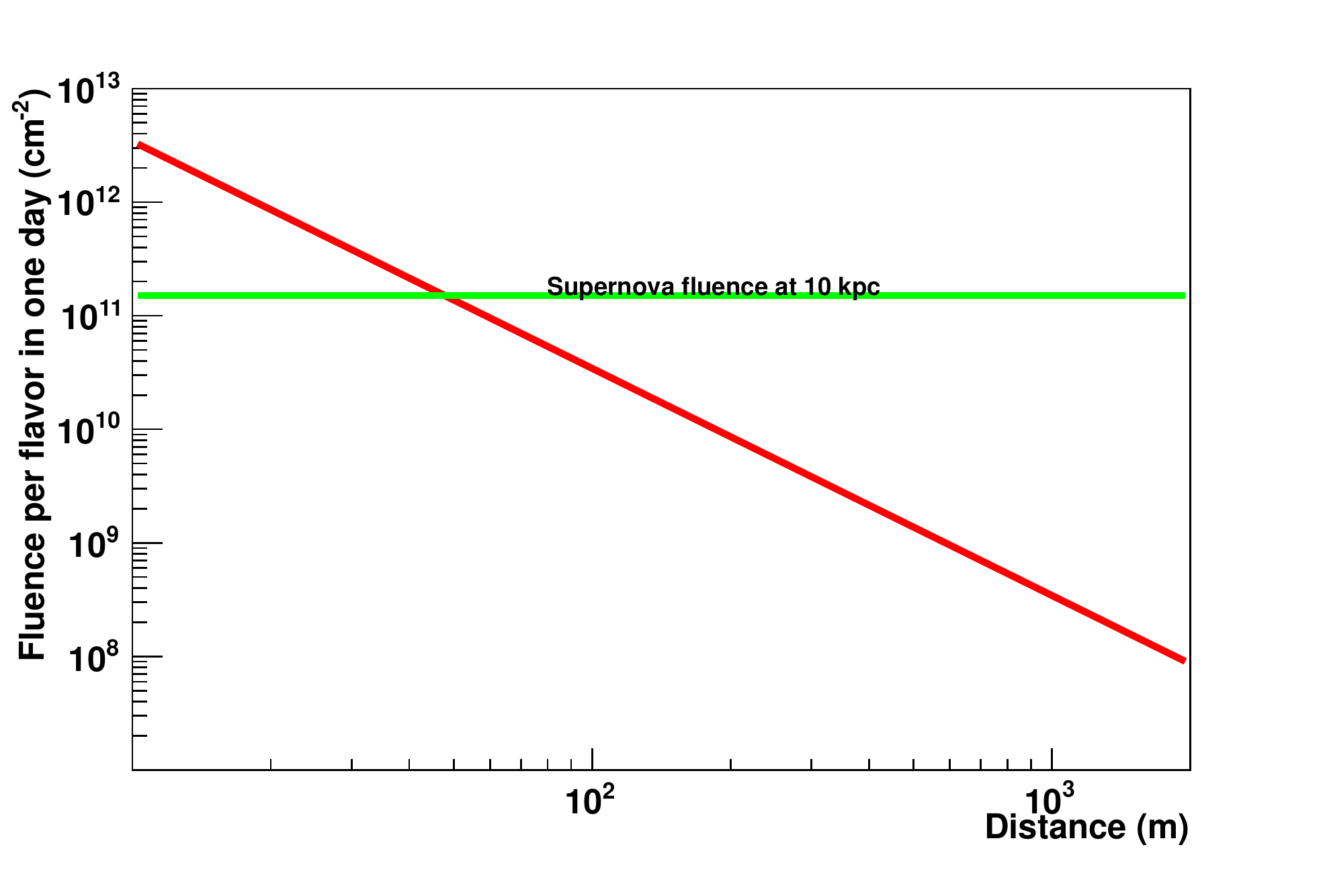}
\vspace{1mm}
\caption{
The red line shows integrated fluence per flavor in one day for the SNS neutrino flux as a function of distance from the source.   
The green solid line shows approximate  fluence per flavor for a supernova at 10 kpc for the full burst. 
}
\label{fig:fluence}
\end{figure}

The time structure of the SNS beam is particularly advantageous for neutrino studies. Time correlations between candidate events and the SNS proton beam pulse will greatly reduce background rates.  As shown in the top left panel of Fig.~\ref{fig:sns_nuplots}, all neutrinos will arrive within several microseconds of the 60-Hz proton beam pulses. As a result, background events resulting from cosmic rays will be suppressed by a factor of $\sim$2000 by ignoring events which occur too long after a beam pulse. At the beginning of the beam spill, the neutrino flux is dominated by muon neutrinos resulting from pion decay, in principle making it possible to isolate pure NC events, since the $\nu_\mu$ in the source have energies below the charged-current (CC) threshold. 
Backgrounds from beam-induced 
high-energy neutrons are present, but can be mitigated by
appropriate siting and shielding.
We note that beam-induced neutron backgrounds
are greatly suppressed for $t\gsim$1~s after the start of the beam spill, while the neutrino production, governed by the muon lifetime ($\tau_\mu \sim$ 2.2 $\mu$s), proceeds for several microseconds. This time structure presents a great advantage over a long-duty-factor facility such as the Los Alamos Neutron Science Center (LANSCE), where the LSND experiment was located~\cite{Athanassopoulos:1996ds}.  Figure~\ref{fig:fluence} shows the expected fluence at the SNS, compared to what would be expected for a nearby supernova.

In general, one wants high neutrino flux (with flux roughly proportional to proton beam power), sharp pulses to enable rejection of off-beam background, and well-understood neutrino spectra.   Ideally pulses should be shorter than than the muon-decay lifetime, and separated by at least several $\tau_\mu$.  Proton energies are ideally less than about 1~GeV in order to minimize contamination from, e.g., kaons and decay-in-flight pions, which can be produced at a significant rate at higher proton beam energies.  Proton energies and target configuration resulting in a high fraction of pion decays at rest will lead to a clean decay-at-rest spectrum and well-known flavor composition.
The SNS satisfies all of these requirements, and
overall is the only facility that within the next decade can provide 1-MW-level intensity, short duty factor, clean decay-at-rest neutrinos.   We note that a second SNS target station may eventually be built; this would provide additional flux, although timing characteristics are as yet unknown.

\section{Physics Motivations}\label{physics}

\subsection{Coherent Elastic Neutrino-Nucleus Scattering}\label{coherent_phys}

The cross section for a spin-zero nucleus, neglecting
radiative corrections, is given by~\cite{Horowitz:2003cz},

\begin{equation}
\frac{d\sigma}{dT}(E,T) = \frac{G_{F}^{2}}{2\pi}M \left[ 2 - \frac{2T}{E} +
\left(\frac{T}{E}\right)^{2} - \frac{MT}{E^{2}}\right]
\frac{Q_{W}^{2}}{4}F^{2}(Q^{2})\,.
\label{eq:dsigmadT}
\end{equation}

where $E$ is the incident neutrino energy, $T$ is the nuclear recoil
energy, $M$ is the nuclear mass, $F$ is the ground-state elastic form
factor, $Q_w$ is the weak nuclear charge, and $G_F$ is the Fermi
constant.  
The condition for coherence requires that momentum transfer
$Q\lsim \frac{1}{R}$,
where $R$ is the nuclear radius. 
This condition is largely satisfied for neutrino energies up to 
$\sim$50~MeV
for medium $A$ nuclei.
Typical values of the total
coherent elastic cross section are in the range $\sim
10^{-39}$~cm$^2$, which is relatively high compared to other neutrino
interactions in this energy range (e.g., CC
inverse beta decay on protons has
a cross section $\sigma_{\bar{\nu}_e p}\sim 10^{-40}$~cm$^2$, and 
elastic
neutrino-electron scattering has a cross section\
$\sigma_{\nu_e e}\sim 10^{-43}$~cm$^2$).

\begin{figure}
\vspace{5mm}
\centering
\includegraphics[width=10cm]{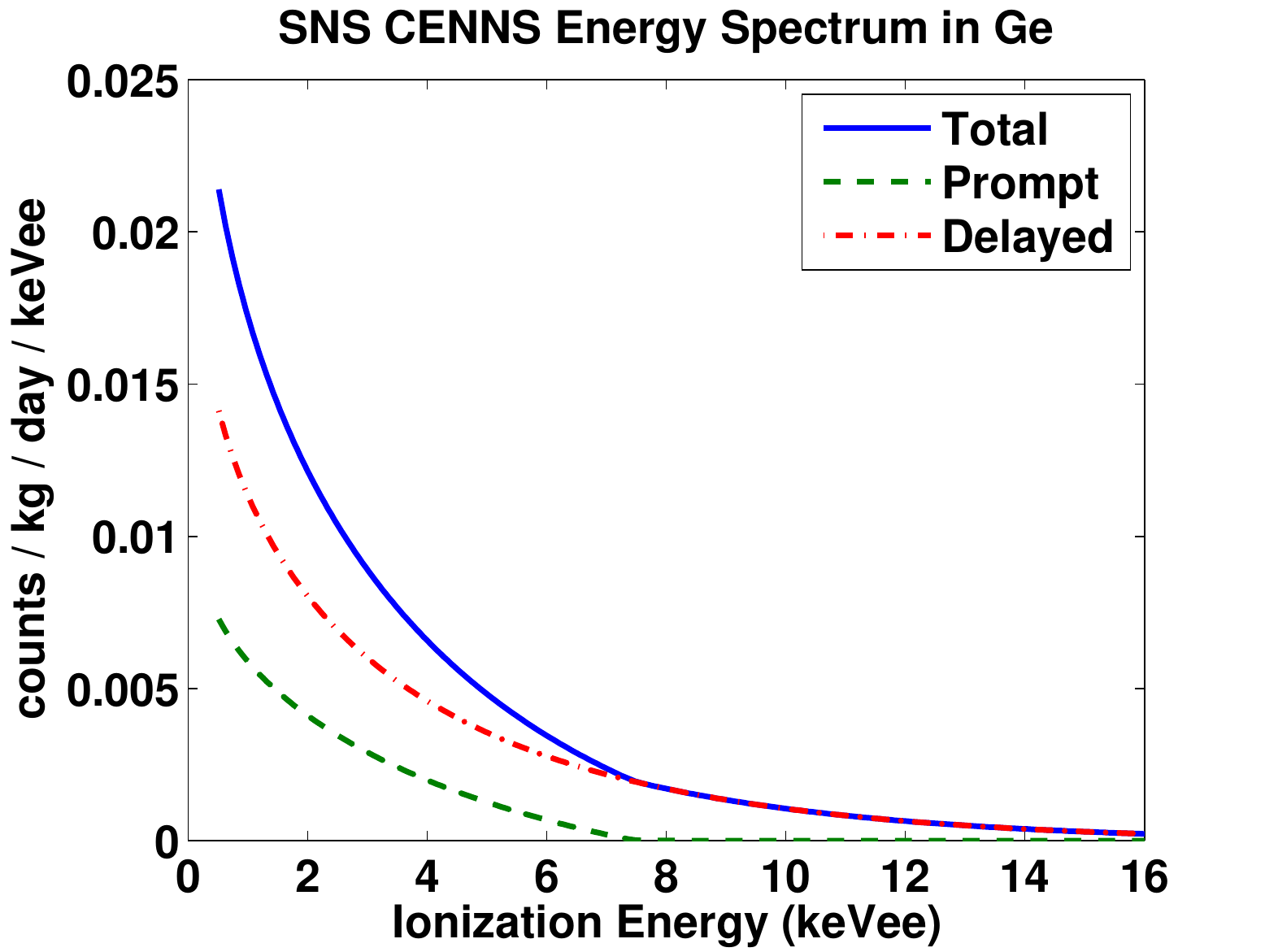}
\vspace{1mm}
\caption{
CENNS ionization-energy (keVee) spectrum (events per keVee per kg active mass of Ge per day of observation time) in a Ge detector located 20 m from the SNS target.  The prompt and delayed components of the spectrum are shown separately.
}
\label{fig:sns_rate_ge_linear}
\end{figure}

Although ongoing efforts to observe CENNS
at reactors~\cite{Barbeau:2007qi,Collar:2008zz,Wong:2005vg}
are
promising, a stopped-pion beam has several advantages with
respect to the reactor experiments. Higher recoil energies bring
detection within reach of the current generation of low-threshold
detectors which are scalable to relatively large target
masses. Furthermore, 
the pulsed nature of the source (see Fig.~\ref{fig:sns_nuplots}) allows
both background reduction and precise characterization of the
remaining background by measurement during the beam-off period.
Finally, the different flavor content ($\nu_e,\nu_\mu,\bar{\nu}_\mu$) of the SNS flux means 
that 
physics sensitivity is complementary to that for reactors, which provide only
$\bar{\nu}_e$; for example, non-standard neutrino interactions (NSI) may be flavor-dependent.
The expected rates for the SNS are quite promising for relevant low-energy detector materials~\cite{Scholberg:2005qs}: see Fig.~\ref{fig:sns_rate_ge_linear}.  

CENNS
reactions are important in stellar core-collapse 
processes~\cite{Freedman:1977xn},
as well as being useful for core-collapse supernova 
neutrino
detection~\cite{Horowitz:2003cz}.  A rate measurement will
have bearing on supernova neutrino physics.
The CENNS cross section is
predicted by the SM, and form-factor
uncertainties are small~\cite{Horowitz:2003cz}.  Therefore a measured
deviation from prediction could be a signature of new physics.
We also note that successful measurement of CENNS in the energy range of solar and atmospheric neutrinos will be immediately useful for direct dark matter search experiments, for which solar and atmospheric neutrinos will eventually represent a background~\cite{Monroe:2007xp,Billard:2013qya}.

We note also that the measured neutrino flux will be a valuable input to the proposed OscSNS~\cite{Elnimr:2013wfa} and CAPTAIN~\cite{Berns:2013usa} experiments, and CENNS could potentially be used for sterile neutrino oscillation searches (e.g.,~\cite{Anderson:2012pn}).

A few of the possible physics measurements are described in more detail below.

\subsection{Standard Model Tests With Coherent Scattering}

According to Eq.~\ref{eq:dsigmadT}, the SM predicts a coherent elastic scattering rate proportional to
$Q_w^2$, the weak charge given by $Q_w = N-(1-4\sin^2 \theta_W)Z$,
where $Z$ is the number of protons, $N$ is the number of neutrons, and
$\theta_W$ is the weak mixing angle.  Therefore the weak mixing angle
can be extracted from the measured absolute cross section, at a
typical $Q$ value of 0.04~GeV/$c$.  
A deviation from the SM prediction
could indicate new physics.  If the absolute cross section can be
measured to 10\%, there will be an uncertainty on $\sin^2 \theta_W$ of
$\sim 5 \%$.  One might improve this uncertainty by looking at ratios
of rates in targets with different $N$ and $Z$, to
cancel common flux uncertainties; future use of enriched neon is a possibility.
There are existing precision
measurements from atomic parity
violation~\cite{Bennett:1999pd,Eidelman:2004wy}, SLAC
E158~\cite{Anthony:2005pm} and NuTeV~\cite{Zeller:2001hh}.
However there is no previous
neutrino scattering measurement in this region of $Q$.
This $Q$ value is relatively
close to that of the Qweak parity-violating
electron scattering experiment at JLAB~\cite{vanOers:2007if,Androic:2013rhu}.
However CENNS tests the SM in a different channel and therefore
is complementary.

In particular, such an experiment  can search for non-standard interactions (NSI)
of neutrinos with nuclei.  Existing and planned precision measurements
of the weak mixing angle at low $Q$ do not constrain new physics that
is specific to neutrino-nucleon interactions.
Reference~\cite{Barranco:2007tz} explores the sensitivity of a 
CENNS experiment on the ton scale to some specific
physics beyond the standard model, including models with extra 
neutral gauge
bosons, leptoquarks and R-parity breaking interactions.

The signature of NSI is a deviation from the expected cross section.
Reference~\cite{Scholberg:2005qs} explores the sensitivity of an
experiment at the SNS.
As shown in the reference, under reasonable assumptions, if the rate
predicted by the SM is observed, neutrino scattering
limits more stringent than
current ones~\cite{Dorenbosch:1986tb, Davidson:2003ha} 
by about an order of magnitude can be obtained for some parameters.

Searches for NSI are based on precise knowledge of the nuclear form factors, which are known to better than 5\%~\cite{Horowitz:2003cz}, so that a deviation from the SM prediction larger than that would indicate
physics beyond the SM.  

Another possibility is to use CENNS to look for anomalous neutrino magnetic moment~\cite{Scholberg:2005qs}, for which the signature would be a distortion of the low-energy recoil spectrum~\cite{Vogel:1989iv}.   The existing limits on magnetic moment for muon neutrinos are relatively weak~\cite{Auerbach:2001wg}, and the muon-flavor content of the stopped-pion neutrino flux would enable improvement of these limits.  Such measurements require very good understanding of detector energy response at low energy.

\subsection{Nuclear Structure from Coherent Scattering}

If we assume that the SM is a good description of nature,  then with sufficient
precision one can measure neutron form factors. 
This physics could
be within reach of a next-generation coherent scattering experiment.
One of the basic properties of a nucleus is its size, or radius, typically defined as
\begin{equation}
\langle R_{n,p}^{2} \rangle^{1/2} = \left( \frac{\int{\rho_{n,p}(r) r^{2} d^{3}r}}{\int{\rho_{n,p}(r) d^{3}r}}\right)^{1/2},
\label{eq:moment}
\end{equation}
where $\rho_{n,p}(r)$ are the neutron and proton density distributions.  Proton distributions in nuclei have been measured in the past to a high degree of precision.  In contrast, neutron distributions are still poorly known.  A measurement of neutron distributions could have an impact on a wide range of fields, from nuclear physics to astrophysics.

Previous measurements of the neutron radius have used hadronic scattering, and result in uncertainties of about $\sim 10\%$ \cite{Horowitz:1999fk}.  A new measurement, being done at Jefferson Laboratory by the PREX experiment, uses parity-violating electron scattering to measure the neutron radius of lead.  The current uncertainty in the neutron radius from this experiment is about $2.5\%$ \cite{Abrahamyan:2012gp}.  An alternate method, first suggested in~\cite{Amanik:2009zz}, is to study the neutron radius through neutrino-nucleus coherent scattering.  
References~\cite{Amanik:2007ce, Patton:2012jr,Patton:2013nwa} explore this possibility.  This measurement is very challenging, as it requires very large statistics and excellent understanding of detector energy response; however it is conceivable for a next-generation CENNS experiment.

\section{Experimental Opportunities}\label{experiments}

\subsection{Detectors to Measure Coherent Elastic $\nu$A Scattering}\label{cohmeas}

We envision approximately three experimental phases on different scales that will address different physics:

\begin{itemize}
\item[] Phase 1: a few to few tens of kg of target material (depending on distance to the source) could make the first measuremenet.
\item[] Phase 2: a few tens to hundreds of kg of target material could set significant limits on NSI, and could also begin to address sterile neutrino oscillations, depending on configuration.
\item[] Phase 3: a ton-scale or more experiment could begin to probe neutron distributions and neutrino magnetic moment.
\end{itemize}

Various technologies are suitable at different scales.   For the first phase, we have settled on
three possibilities for prompt deployment:  CsI[Na], germanium PPCs, and two-phase xenon.  Conceivably more than one technology could be deployed simultaneously, depending on available resources, siting and background issues.

The following subsections describe these different possibilities.

\subsubsection{CsI[Na] Detectors}

\begin{figure}
\includegraphics[width=18.cm]{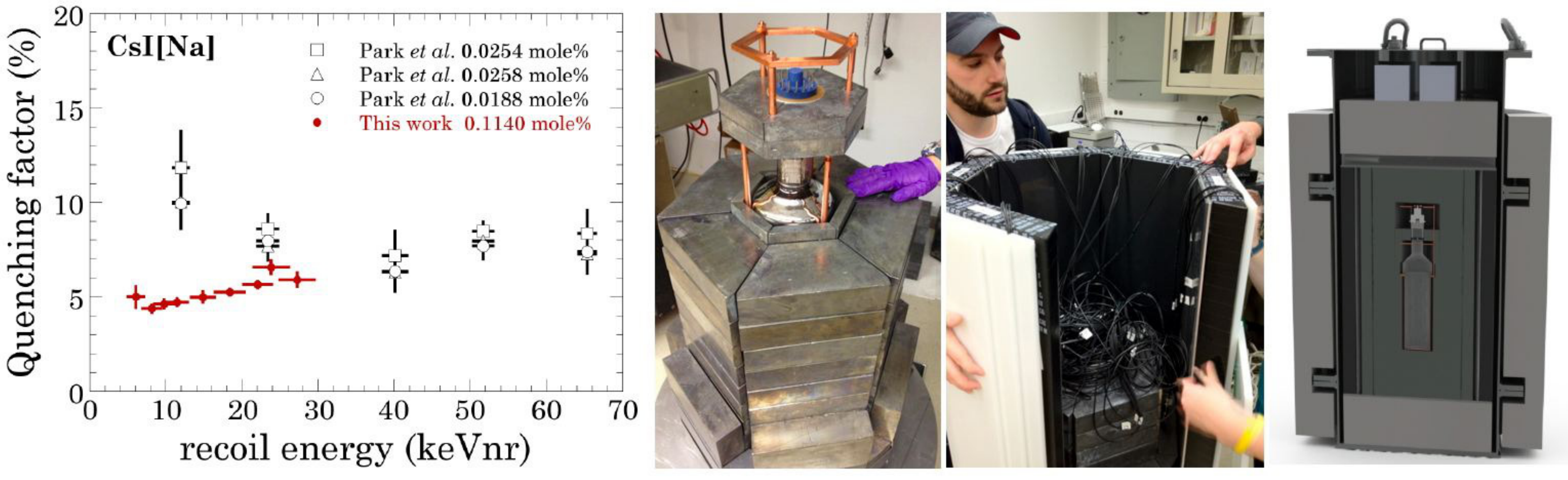}
\caption{\label{fig:epsart}  {\it Left:} Quenching factor for low-energy nuclear recoils in CsI[Na], measured using the method described in \protect\cite{Collar:2013gu}. The range of recoil energies is relevant for SNS neutrino energies. {\it Center panels:} Assembly of a 2 kg CsI[Na] prototype within the shield intended for SNS installation. The detector features selected salts screened against radioactive contaminants, OFHC copper parts, a low-background ET 9390UFL photomultiplier, archeological-quality lead shielding ($<$0.02 Bq/kg $^{210}$Pb), a 99.6\% efficient muon veto, and neutron moderator and absorber. {\it Right:} Full model of the shield containing the 15 kg detector, showing the steel container used to lower the assembly 
into a shallow pit.}\label{fig:cosi}
\end{figure}

CsI[Na] scintillators present several advantages for CENNS neutrino measurements. These are briefly listed below:

\begin{itemize}

\item The large CENNS cross-section from both recoiling species, cesium and iodine, provides $\sim$800 recoils per 15 kg of CsI[Na] per year above the expected $\sim$5 keVnr threshold of this detector. Both recoiling species are essentially indistinguishable due to their very similar mass, greatly simplifying understanding the response of the detector. 

\item The quenching factor for nuclear recoils in this material over the energy region of interest has been carefully characterized, using the methods described in \cite{Collar:2013gu} (Fig.~\ref{fig:cosi}). Its value is sufficiently large to expect a realistic $\sim$5 keVnr threshold. 

\item It should be possible to perform statistical discrimination (as opposed to event-by-event discrimination) between nuclear and electron recoils at the level of $\sim$1,000 accumulated events. This is based on a difference of $\sim$60 ns between the scintillation decay times observed for these two families of events \cite{cosinima}. Similar differences have already been exploited to implement this discrimination in NaI[Tl] scintillators dedicated to WIMP detection. 

\item Prototype crystals grown from screened salts containing low levels of U, Th, $^{40}$K, $^{87}$Rb, and $^{134,137}$Cs should provide a neutrino signal-to-background ratio of $\mathcal{O}$(5), even prior to (anticoincident) background subtraction.

\item Several other practical advantages exist: CsI[Na] exhibits a high light yield (64 photons/keVee) and has the best match to the response curve of bialkali photomultipliers of any scintillator. It is a rugged room-temperature detector material, and is also relatively inexpensive ($\sim$ \$1/g), permitting an eventual increase in mass to a $\sim$100 kg target,  which would allow to explore the most interesting physics goals planned for CSI. CsI[Na] lacks the excessive afterglow (phosphorescence) that is characteristic of CsI[Tl] \cite{cosinima}, an important feature in a search involving small scintillation signals in a detector operated at ground level.

\end{itemize}

A 15 kg CsI[Na] detector (CoSI) and its shielding (Fig.\ 1), built by University of Chicago and Pacific Northwest National Laboratory CSI collaborators, is being completed at the time of this writing, aiming at installation at the SNS during 2013. This device should be sufficient for a first measurement of CENNS at the SNS, while providing operational experience towards a final array of crystals with total mass $\mathcal{O}$(100) kg. 

\subsubsection{Germanium Detectors}

A powerful approach to detecting CENNS at the SNS is to use point-contact high-purity germanium (HPGe) detectors.  HPGe technology is well matched to the problem of detecting nuclei recoiling from CENNS interactions.   For a 5-keV nuclear recoil, the quenching factor (conversion from nuclear recoil energy to equivalent ionization energy) is about 20\%, so the signal is about 1-keV ionization energy.  Point contact HPGe detectors have very small electrodes, and so have very small detector capacitance, and consequently very low electronic noise \cite{Luke:1989}.  When cooled, the leakage currents can also be very low
--less than 1 pA-- so the current noise is also small.  P-type point contact (ppc) detectors with an electronic noise FWHM of order 160~eV have been demonstrated \cite{Aalseth:2011wp,Aalseth:2012if}, and so should be able to run with detection thresholds below 1 keV.   The low noise leads to excellent energy resolution, so the observed (background subtracted) energy-deposition spectrum is close to the actual spectrum. 

PPC detectors have a relatively long drift time; it can be more than 1 $\mu$s \cite{Martin:2011vj}.  However, the detector timing will still be sufficient to reject out-of-time backgrounds, and allow for at least statistical separation of the prompt and delayed neutrino components.  

The high intrinsic radio-purity of the Ge itself and the availability of low-background cryostats make possible low-background operation.   Existing detectors have demonstrated intrinsic background levels suitable for the CENNS measurement \cite{Aalseth:2012if}.

Ge detectors are a mature technology. HPGe detectors are widely used, and even point-contact detectors are commercially available from, e.g., CANBERRA and ORTEC.

Figure 1 shows the spectrum of energy deposition, in terms of equivalent ionization energy (keVee) for a detector located 20 m from the SNS target, where the neutrino flux is  $2\times 10^7\nu/\text{s}/\text{cm}^2$.  The prompt and delayed components to the spectrum are shown separately.   The cross-section form factor is modeled following \cite{Klein:1999qj} and the quenching factor in Ge following \cite{Barbeau:2009zz}.   

At a distance of 20 m from the target, for a 1 keV ionization energy threshold, the signal is approximately 3.6 events/kg/month.  A detector with 5 kg of active mass would collect 100 signal events in about 6 months.    Most current ppc detectors, e.g., the {\sc Majorana Demonstrator}, have masses of order 0.5-1.1 kg~\cite{Elliott:2013eqb}.
An experiment using an array of five crystals in a single cryostat would thus have the required sensitivity.

By using a low-background detector and shielding, and taking advantage of the pulsed structure of the SNS beam, the intrinsic detector backgrounds would be far below the expected signal rate.  Because germanium is relatively dense (5.3~g/cm$^3$), the detector will be compact, and a fairly simple scintillator-based active veto could reduce the cosmic-ray muon rate to far below background.  Cosmic-spallation neutrons and beam-related neutrons are left as the challenging sources of background.  Efforts are underway to measure these neutron backgrounds and determine how a shielded germanium ppc detector responds to them. 

In the longer term, the excellent germanium energy resolution should also lend itself to more precise studies of CENNS, by adding more detectors to the array.

\subsubsection{Two-Phase Xenon Detectors}

Since 1992, the employment of a liquid xenon detector has been considered 
for neutrino magnetic moment searches by probing $\bar{\nu}_e$-electron scattering cross sections for deposited energies below 100 keV.
Such an experiment could be performed with a moderately-sized ($\sim$1 ton mass)  LXe emission detector~\cite{BaldoCeolin:1992yj}.

\begin{figure}
\vspace{5mm}
\centering
\includegraphics[width=7.5cm]{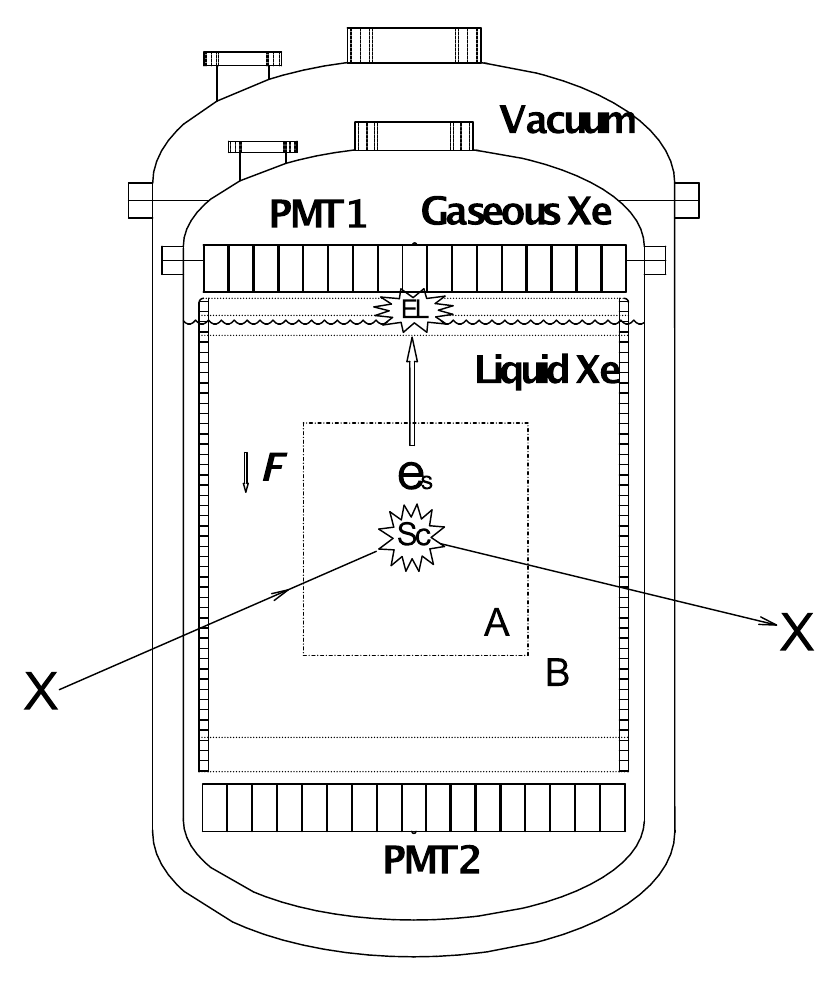}
\vspace{1mm}
\caption{
Principle of operation for a “wall-less” liquid xenon emission detector detecting a hypothetical weakly-interacting particle X. Sc:  scintillation flash generated at the point of primary interaction between X and Xe atoms; EL: electroluminescence flash of gaseous Xe excited by electrons extracted from liquid Xe by electric field F and drifting through the gas at high electric fields ($>$1 kV/cm/bar); PMT1 and PMT2: arrays of photodetectors detecting Sc and EL signals; A: the fiducial volume where events considered to be useful occur; B: the shielding layer of LXe. The active volume of the detector is surrounded with highly reflective cylindrical PTFE reflector embodied with drift electrode structure providing a uniform field F. The detector is enclosed in a vacuum cryostat made of low-background pure titanium.
}
\label{fig:xenon1}
\end{figure}

The emission method of particle detection invented 40~years ago at the Moscow Engineering Physics Institute (MEPhI) Department of Experimental Nuclear Physics~\cite{Dolgoshein:1970} allows an arrangement of a ``wall-less'' detector sensitive to single ionization electrons~\cite{Bolozdynya:1995}. 
See Fig.~\ref{fig:xenon1}.    A detector using this technology is also sensitive to CENNS-induced recoils.

To measure the ionization yield of heavy nuclear recoils, an experiment is underway that will model the detection of xenon nuclear recoils through a study of the elastic scattering of a monochromatic filtered beam of neutrons from the IRT MEPhI 
reactor with a 5-kg LXe emission detector recently used for detection of single electrons~\cite{enpl,Solovov:2011mna,Akimov:2012zz,Akimov:2013gla}. 

For observation of CENNS at the SNS, we consider the emission detector RED100 with a 100-kg LXe working medium. The basic parameters for construction of this detector are the following:
a low-background titanium cryostat, a readout system based on two arrays of low-background PMTs (e.g., Hamamatsu R11410) located in the gas phase and in the liquid below the grid cathode, and a PTFE light-collection system embodied with a  drift electrode system as shown in Fig.~\ref{fig:xenon1},
a cryogenic system based on thermo-siphon technology similar to that used in the LUX detector, and
a location in a borehole 10~m underground at 40~m distance from the target.
For these conditions, 
the total expected event rate is 1470 events/year.

\subsubsection{Other Possibilities: Argon/Neon Single-Phase Detectors}

\begin{figure}[!ht]
  \centering
    \includegraphics[height=2.0in]{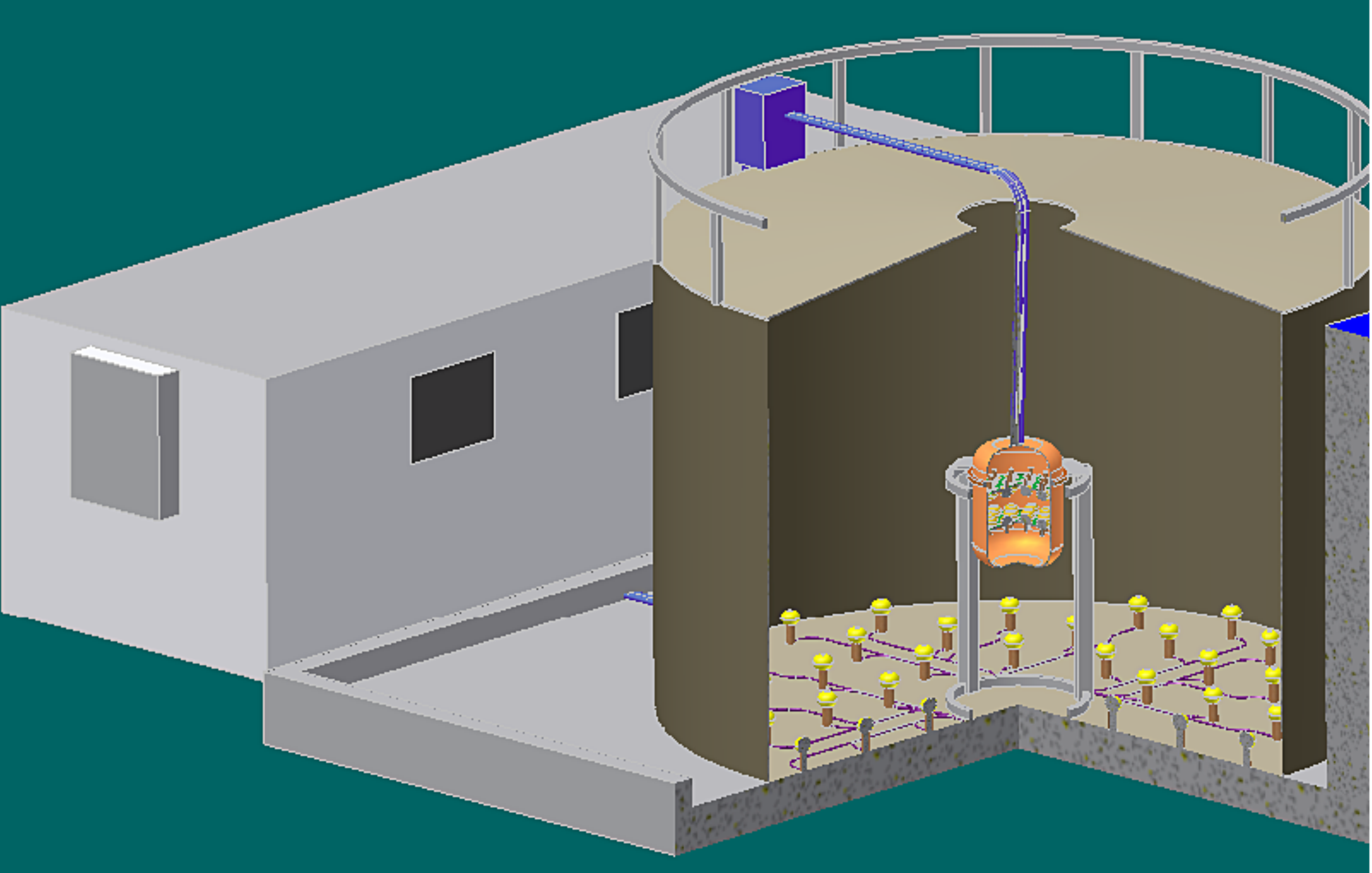}
  \caption{CLEAR experiment concept.  The cryogenic inner
    detector enclosed in a vacuum vessel is positioned inside a tank
    of water, which provides neutron shielding and an active muon veto
    by detection of Cherenkov radiation with an array of PMTs.
   }\label{fig:overview}
\end{figure}

For a Phase II experiment, single-phase argon/neon is a possibility~\cite{Scholberg:2005qs}. Such a noble-liquid neutrino detector is conceptually similar to dark-matter detectors using liquid argon. This kind of detector will utilize pulse-shape discrimination of scintillation light between nuclear recoil and electron recoil interactions (and ionization yield) in the liquid argon to identify CENNS interactions from  background events. The majority of electromagnetic and neutron backgrounds can be rejected using the standard active and passive shielding methods together with self-shielding fiducialization. 
A specific detector proposed to accomplish these goals was designed,
called CLEAR (Coherent Low
Energy A (Nuclear) Recoils~\cite{Scholberg:2009ha}).  This concept employed a single-phase design to allow
interchangeable noble liquid target materials, which is advantageous because multiple targets are desirable to test
for physics beyond the SM..
This design comprises an inner noble-liquid detector
placed inside a water tank. The water tank instrumented with
PMTs acts as a cosmic ray veto. An overview diagram
of the experiment is shown in Fig.~\ref{fig:overview}.

\subsection{Siting Possibilities}

\begin{figure}
\vspace{5mm}
\centering
\includegraphics[width=9cm]{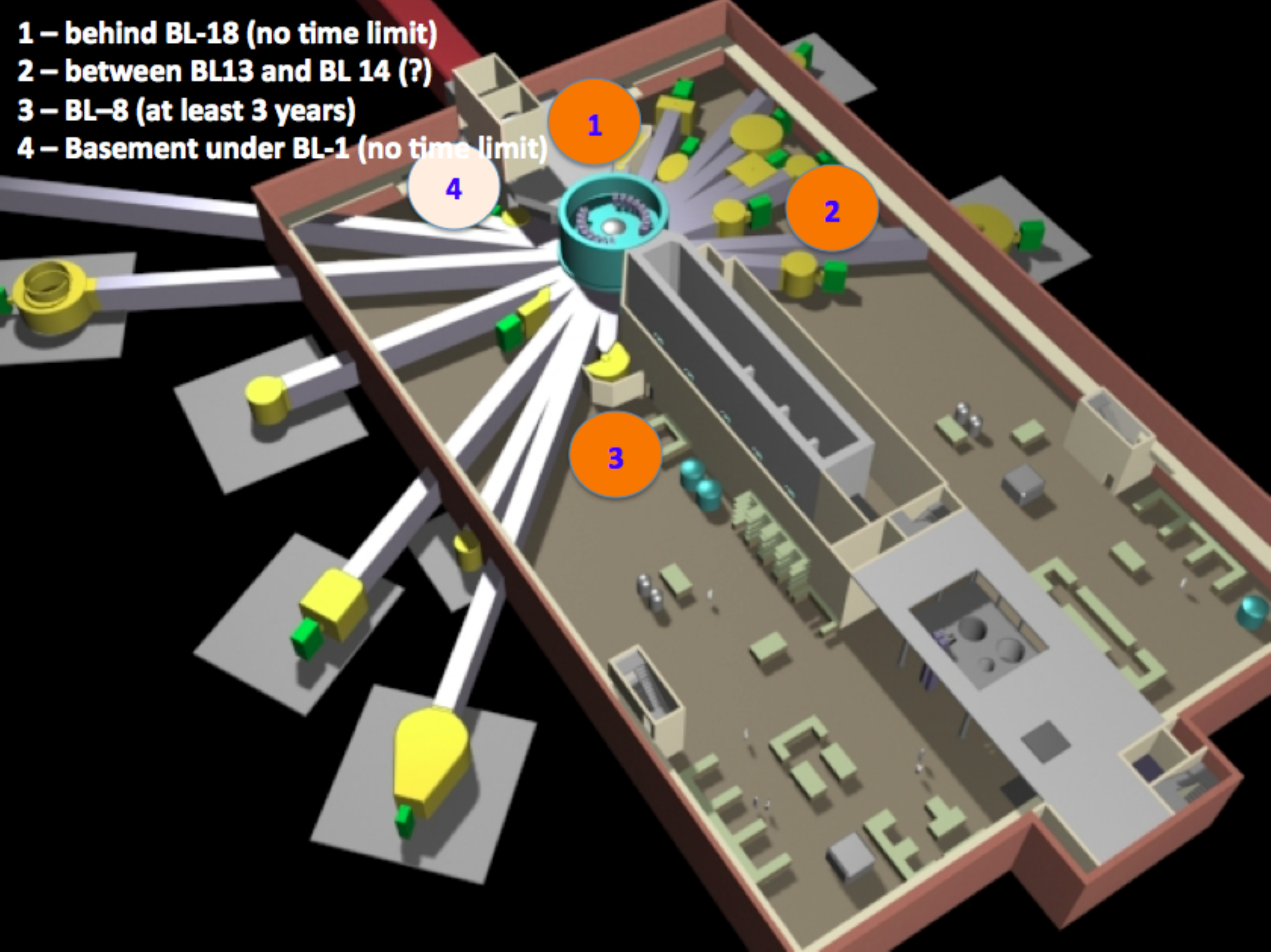}
\vspace{1mm}
\caption{Potential sites for deployment of a CENNS experiment at the SNS.}
\label{fig:sites}
\end{figure}

Several potential sites have been identified inside the target building at which a CENNS experiment could be deployed within the next few years. They are shown in Fig.~\ref{fig:sites}.   One of these (site 4) is in a basement with a few meters of extra overburden; the others are on the SNS target building floor level.  At these locations, detectors can be deployed at distances between around 15-35 meters from the SNS source.
Additional shielding can be deployed at these locations.

A site outside the SNS target building is also possible, although less desirable due to farther distance from the neutrino source.  A detector could potentially be deployed in a borehole for shielding.   However it will be more expensive to deploy an experiment outside the SNS target building due to extra construction costs.

A siting decision will be made pending results of the background studies described in the next section.

\begin{figure}
\vspace{5mm}
\centering
\includegraphics[width=7cm]{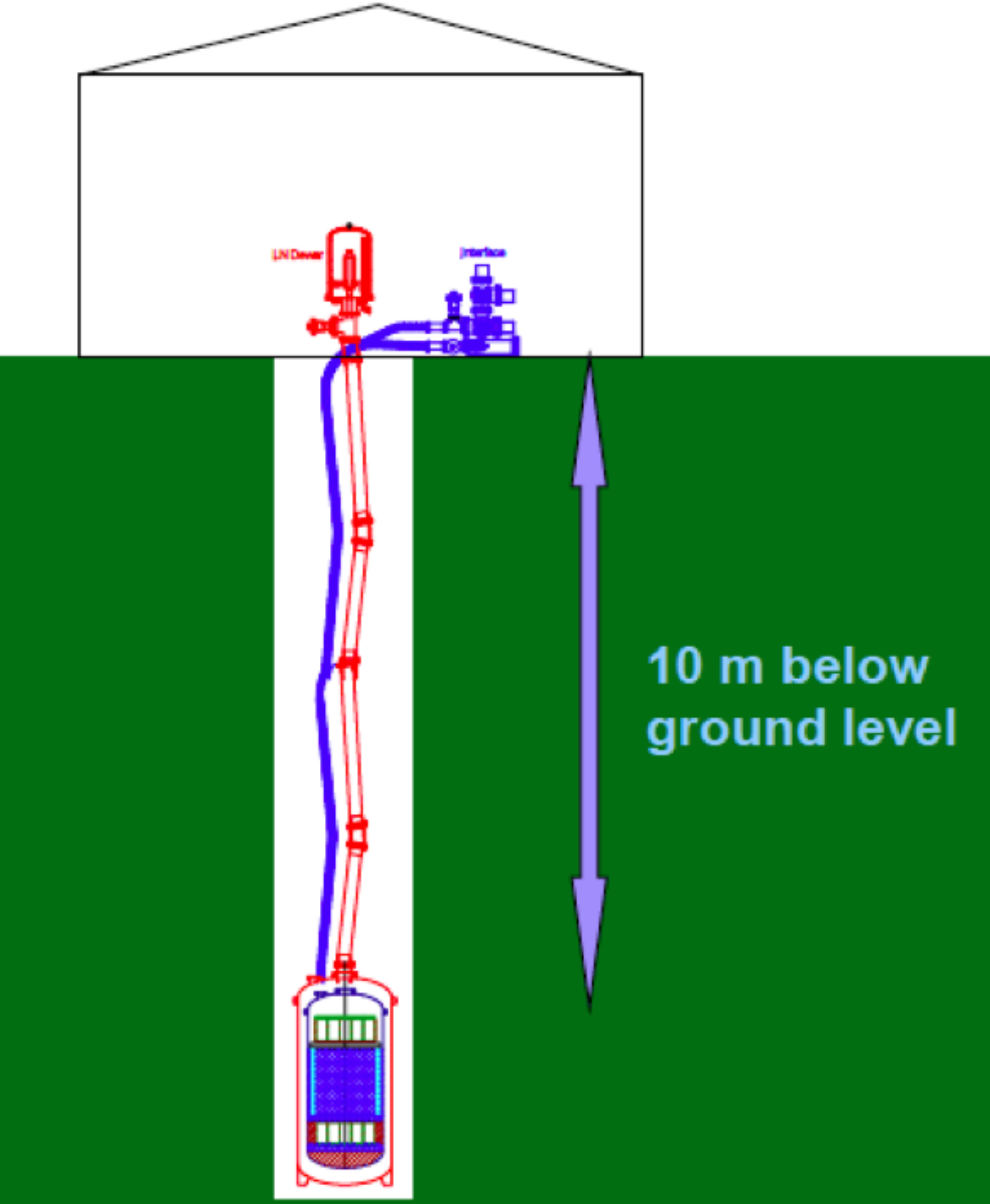}
\vspace{1mm}
\caption{
Borehole concept for detector deployment.
}
\label{fig:hole}
\end{figure}

\subsection{Backgrounds}

Understanding and reduction of backgrounds are critical for the CSI experimental program.  Steady-state backgrounds, such as radioactivity and cosmogenics, will be reduced by the SNS duty factor ($10^3-10^4$), and may or may not need additional mitigation with shielding and cleanliness of detector materials, depending on siting and detector properties.  Steady-state backgrounds can also be very well understood using data taken outside the SNS beam window.
However, beam-related backgrounds, especially fast neutrons, will have to be very carefully studied, using ancillary measurements and modeling, and shielded against.  These beam-related backgrounds will likely be highly site-dependent, and also possibly time-dependent.

Currently a background measurement campaign is underway at the SNS to characterize the backgrounds at the candidate locations and during different SNS running conditions.  These measurements are useful also for SNS neutron experiments.
Three instruments are taking data: a neutron scatter camera, a liquid scintillator array, and a low-background point-contact germanium detector.  More detectors may be deployed in the near future.

\section{Conclusion}

We have outlined a program for taking advantage of the extremely high-quality stopped-pion neutrino source available at the Spallation Neutron Source available at Oak Ridge National Laboratory for CENNS measurements.  Work is currently underway to evaluate backgrounds and siting options.

\section*{Acknowledgements}

We are grateful for excellent logistical support and advice from local SNS and Oak Ridge personnel.  

\bibliographystyle{vitae}
\bibliography{refs}

\end{document}